\documentclass[runningheads]{llncs}
\usepackage[T1]{fontenc}
\usepackage{amsmath}
\usepackage{graphicx}
\usepackage{cite}
\usepackage{url}
\usepackage{hyperref}

\begin{document}
\title{GATES: Graph Attention Network with Global Expression Fusion for Deciphering Spatial Transcriptome Architectures%
}

\titlerunning{GATES}
% If the paper title is too long for the running head, you can set
% an abbreviated paper title here
%

\author{Xiongtao Xiao\inst{1}\and 
Xiaofeng Chen\inst{1} \and
Feiyan Jiang\inst{1} \and
Songming Zhang\inst{2} \and
Wenming Cao\inst{1}\and
Cheng Tan\inst{3}\and
Zhangyang Gao\inst{3}\and
Zhongshan Li\inst{4}}
\authorrunning{X. Xiao et al.}
\institute{Chongqing Jiaotong University, Chongqing, China\\
\email{xiaoxxtt@mails.cqjtu.edu.cn, matcxf@cqjtu.edu.com, hhfyjiang@mails.cqjtu.edu.cn, wmingcao@cqjtu.edu.cn}
\and
Shenzhen University of Advanced Technology, Shenzhen, China\\
\email{sm.zhang1@siat.ac.cn}
\and
Westlake University, Hangzhou, Zhejiang, China\\
\email{tancheng@westlake.edu.cn, gaozhangyang@westlake.edu.cn}
\and
Georgia State University, Atlanta, Georgia, USA\\
\email{zli@gsu.edu}
}
\maketitle              % typeset the header of the contribution
\begin{abstract}
    
Single-cell spatial transcriptomics (ST) offers a unique approach to measuring gene expression profiles and spatial cell locations simultaneously. However, most existing ST methods assume that cells in closer spatial proximity exhibit more similar gene expression patterns. Such assumption typically results in graph structures that prioritize local spatial information while overlooking global patterns, limiting the ability to fully capture the broader structural features of biological tissues. To overcome this limitation, we propose GATES (Graph Attention neTwork with global Expression fuSion), a novel model designed to capture structural details in spatial transcriptomics data. GATES first constructs an expression graph that integrates local and global information by leveraging both spatial proximity and gene expression similarity. The model then employs an autoencoder with adaptive attention to assign proper weights for neighboring nodes, enhancing its capability of feature extraction. By fusing features of both the spatial and expression graphs, GATES effectively balances spatial context with gene expression data. Experimental results across multiple datasets demonstrate that GATES significantly outperforms existing methods in identifying spatial domains, highlighting its potential for analyzing complex biological tissues. Our code can be accessed on GitHub at \url{https://github.com/xiaoxiongtao/GATES}.
    \keywords{Spatial transcriptomics  \and Graph neural network \and Spatial domain identification.}
\end{abstract}

\section{Introduction}
Spatial transcriptomics (ST) technology~\cite{advances, introduction, expanding} allows gene expression levels of single or several cells to be studied at different spatial locations. How to understand gene expression patterns correlate with the spatial context of the tissue microenvironment is a complex issue~\cite{museum}. Biological tissues are composed of different types of cells (muscle cells, nerve cells, $etc$), and the structural characteristics of such tissues include the relative position of cells (tight junction, desmosomes junction, free state, $etc$) and the spatial distribution of gene expression. These elements constitute a multifaceted organizational framework. Therefore, identifying spatial domains~\cite{identifying, Giotto,stamarker, BayesSpace} in ST data is critical to elucidating tissue functions and understanding disease mechanisms.

Clustering algorithms, widely used in unsupervised machine learning, group data based on inherent similarities without explicit labels. 
These methods can be categorized into non-spatial and spatial clustering approaches. Non-spatial methods, such as k-means~\cite{kmean} and Louvain~\cite{louvain}, utilize gene expression data alone to cluster ST data. However, they face significant challenges due to the high dimensionality and sparsity nature of ST data, often resulting in clustering outcomes that do not correspond well to tissue sections and fail to accurately represent spatial domains. In response, researchers have developed spatial clustering methods that leverage spatial information to enhance identifying biological structures. For instance, Giotto~\cite{Giotto} employs a Hidden Markov Random Field (HMRF) model to estimate spatial domain distributions by maximizing the likelihood of observed data. Similarly, BayesSpace~\cite{BayesSpace} improves the resolution of ST data, enabling the detection of subtle tissue structures. Despite these advancements, both methods primarily rely on linear principal component analysis for feature extraction, which restricts their capacity to model complex nonlinear interactions inherent in biological data~\cite{DeepST}.

Up to now, graph neural networks have been used for analyzing ST data. The composition of ST data is used to learn low-dimensional embedding. For example, PROST~\cite{PROST}, STitch3D~\cite{STitch3D}, stMDA~\cite{stMDA}, GraphST~\cite{GraphST}, $etc$, use the spatial information of spots to construct neighbor graphs, and combine them with gene expression profiles as input to graph neural networks. However, these methods overlook the impact of neighboring spots on the gene expression of a given spot. AVGN~\cite{AVGN} and DeepST~\cite{DeepST} used enhanced gene expression data based on the similarity of neighbor spots and combined with spot coordinates to realize the construction of adjacency matrices. STAGATE~\cite{STAGATE} constructed spatial neighbor graphs and utilized a graph attention autoencoder framework to identify spatial domains based on the pre-clustering of gene expressions neighbor graphs. 
The aforementioned methods construct graph structures solely based on the premise that spatially neighboring cells share more similar gene expression profiles. This leads graph networks to prioritize local spatial information, while failing to capture global context, ultimately restricting the network’s ability to identify broader biological structural characteristics.

In this study, we present GATES (Graph Attention neTwork with global Expression fuSion), a novel approach designed to capture both local and global information for precise domain identification in ST data. GATES begins by constructing an expression graph that combines local spatial proximity with global gene expression similarity. It then utilizes an autoencoder with adaptive attention to assign optimal weights to neighboring nodes, enhancing the network's ability to extract relevant features. By linearly combining features from spatial and expression graphs, GATES effectively balances spatial context with gene expression data. Demonstrating consistently high accuracy across datasets from various platforms, GATES proves to be both robust and adaptable. This strong performance underscores its potential to uncover biologically meaningful patterns and insights within diverse spatial transcriptomics technologies.
\section{Results}
\subsection{Overview of GATES}
\begin{figure}[!htb]
\centering
\includegraphics[width=\textwidth]{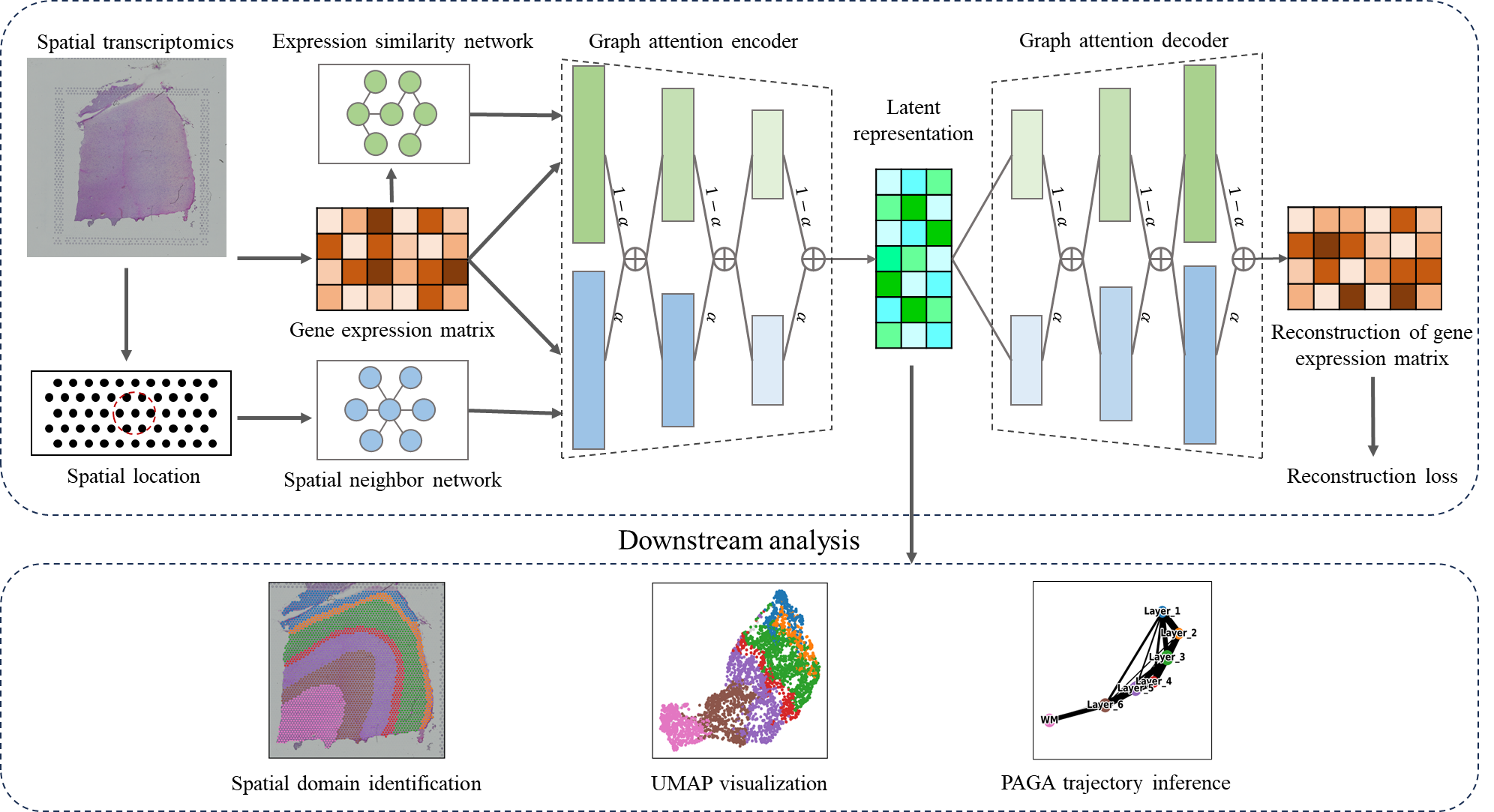}
\caption{Overview of GATES workflow. It uses spatial transcriptomic data (including gene expression and spatial coordinates) of tissue sections to construct the graph of 
spatial neighborhood graph and gene expression similarity graph. Then, these two graphs are utilized as inputs for the encoder to extract feature embeddings and decode to reconstruct the original gene expression profile. The latent representation after training can be used for downstream analysis.} \label{structure}
\end{figure}

GATES is a novel unsupervised learning framework that integrates spatial neighbor graph and gene expression similarity graph, as shown in Fig. \ref{structure}. Specifically, firstly, the spatial neighbor graph is constructed according to the spatial location information of the spot. It is based on the assumption that spatially close spots tend to have more similar gene expression profiles~\cite{BayesSpace,stlearn,unsupervised}. This means that local information about the tissue is taken into account. Meantime, the cosine similarity of each spot was calculated according to gene expression to construct a genetically similar neighbor graph. This diagram indicates that a spot located far from the center can still be considered an adjacent spot. For instance, cells in different locations may exhibit similar gene expression patterns due to shared biological processes or signaling pathways. This means that the global information is taken into account. Thus this graph can reveal biological patterns that are difficult to detect spatially, such as the co-expression patterns of genes in specific spatial domains. While it may not be evident in single-cell or local analyses, it can be more readily identified through global analysis. Then, the two graphs are used as the input of the graph attention autoencoder~\cite{auto-encoder}, and the low-dimensional feature vectors with spatial information and gene expression are extracted through the encoder. 

In order to better integrate the two graph structures, we introduce an attention mechanism between the encoder and the decoder. The attention mechanism is used to give special focus on the low-dimensional embedding feature vectors of the two graphs. Then they are linearly combined according to a certain weight value. And this also significantly enhances the model's ability to capture the complex interactions between these two graph structures. It is decoded and reversed back to the original feature space to reconstruct the gene expression matrix.
Finally, the learned low-dimensional embeddings can be used for a series of downstream tasks, such as spatial domain recognition, UMAP visualization and trajectory inference, to reveal cell heterogeneity and new cell types, and to understand tissue development and disease mechanisms.

\subsection{GATES is robust to the spatial domain recognition of different datasets on different platforms}
\subsubsection{GATES is robust in identifying spatial domains on DLPEC datasets compared to other methods.}
We applied it to the DLPFC ST dataset~\cite{DLPFC} generated by 10x Visium~\cite{10XVisium}. In order to verify the method to improve the accuracy of spatial domain recognition, we compared it with nine detection methods (SpaceFlow~\cite{SpaceFlow}, SpaGCN~\cite{SpaGCN}, Seurat~\cite{SEURAT}, SEDR~\cite{SEDR},ConST~\cite{const}. $etc$.). The spatial domain recognition performance was evaluated by using artificial annotation as the true value. The spatial domain similarity between the predicted and manual annotation was quantified by the adjusted Rand index (ARI). Relevant details are provided in Appendix A.

First, we show a visualization of the spatial domain of some slices on the human dorsolateral prefrontal cortex (DLPFC) ST dataset (Fig. \ref{ABCDEFG1}A and Figs. \ref{domain1}-\ref{domain3} in Appendix B). Through comparison, we found the GATES method has a better effect than the STAGATE for the spatial domain separation of each slice. Especially in the 151675, it can be seen that GATES divides each layer. On the 151673 slices of ground truth, although the fourth region is narrower, the spatial domains identified by our method are exactly the same as the corresponding layers of Ground Truth. However, the STAGATE method identifies a wider effect. This indicates our method exhibits greater precision in identifying narrow regions. In contrast, the STAGATE method has an excessively wide spatial domain identification when dealing with the same region. This may mean that it has certain limitations in spatial resolution, especially when dealing with complex or detailed spatial structures. More surprisingly, STAGATE failed to fully identify the second layer of 151675 and the third layer of 151676. A partial fracture was observed, whereas our method consistently identified each layer (Fig. \ref{ABCDEFG1}A). 
In the embedded UMAP space, the 151673 slice from STAGATE exhibits insignificant separation of spots across different layers, with numerous outliers present at the intervals between layers. Conversely, the results identified by our method show that spots from the same manually annotated layer cluster together, resulting in clearer separation between layers. Notably, in the 151669 slice of the PAGA plot, our method exhibits the developmental trajectories between layers and the similarities among adjacent layers with clarity. It forms a coherent and readily understandable spatial structure. In contrast, the PAGA results embedded by the STAGATE method appear relatively mixed. The similarities and developmental trajectories among adjacent layers are blurred, making it challenging to directly obtain clear spatial domain information(Fig. \ref{ABCDEFG1}B, and Figs. \ref{PAGA1}-\ref{PAGA3} in Appendix B). This further validates the superiority of our method in analyzing complex spatial structures. It facilitates a deeper understanding of cellular developmental processes and differentiation pathways within organisms. By examining the correlations between different layers, we can clearly observe how cells gradually differentiate from their primitive state into various cell types with specific functions.
Furthermore, GATES achieved the highest median ARI scores among the other nine methods when clustering across the 12 slices. In the boxplots for our method and STAGATE, both of which exhibited the highest ARI values, the lower line for the GATES method is significantly shorter than that of STAGATE. This observation indicates minimal variance in ARI values for the identification of spatial domains across the 12 slices. Such consistency suggests that GATES demonstrates exceptional stability and reliability, allowing it to effectively adapt to various types of slice data when identifying complex organizational structures (Fig. \ref{ABCDEFG1}C).

\begin{figure}
\centering
\includegraphics[width=\textwidth]{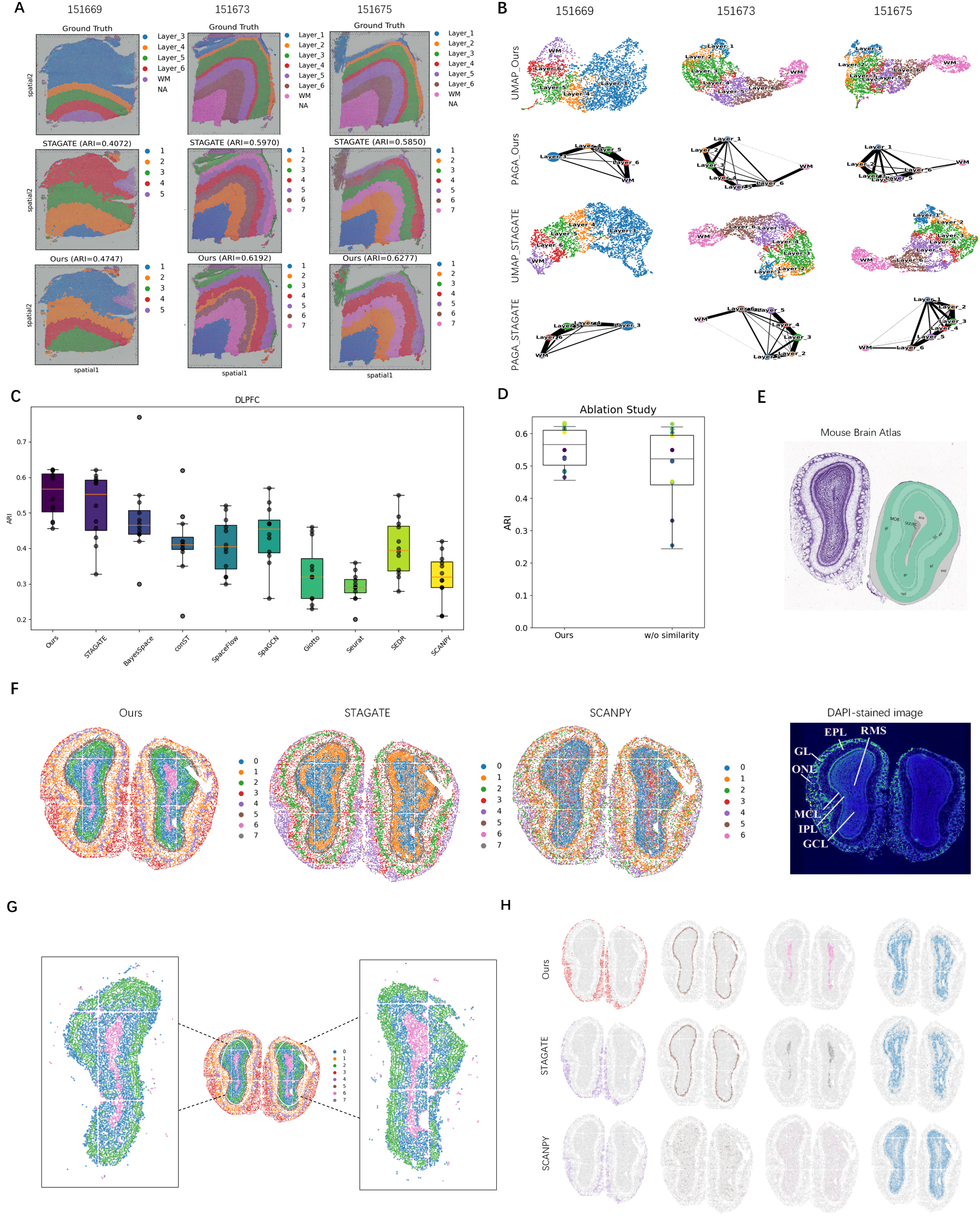}
\caption{the DLPFC and mouse olfactory bulb tissues dataset are used to robustly identify the spatial domain. \textbf{A}. Spatial domain identification by STAGATE and GATES. \textbf{B}. UMAP visualizations (top) and PAGA graphs (bottom) generated by STAGATE and GATES, respectively. \textbf{C}. Spatial domain identification results on 12 slices of DLPFC by different methods. \textbf{D}. GATES retains or removes the spatial domain identification results of the gene expression similarity graph. \textbf{E}. DAPI-stained image (bottom) and the Allen Mouse Brain Atlas (top). \textbf{F}. Visualization of the spatial domain of the mouse olfactory bulb identified by the Gates, STAGATE, and SCANPY. \textbf{G}. The detailed diagram for spatial domain identification using GATES. \textbf{H}. Detailed diagram of the spatial domain of the different methods.} \label{ABCDEFG1}
\end{figure}

\subsubsection{GATES captures the complex spatial structure of the mouse olfactory bulb in greater detail.}
We validated the performance of the GATES method under the mouse olfactory bulb ST dataset generated by the Stereo-seq platform. Normal olfactory bulbs are generally composed of~\cite{STAGATE} the rostral migratory stream (RMS), granule cell layer (GCL), internal plexiform layer (IPL), mitral cell layer (MCL), external plexiform layer (EPL) and olfactory nerve layer (ONL) (Fig. \ref{ABCDEFG1}E). We observed that STAGATE failed to recover the laminar organization, and SCANPY identified some mixture spatial clusters with unclear tissue structures. By contrast, our method shows a clearer delineation of the boundaries between different domains within laminar tissues and aligns well with the annotated laminar structures in the Allen Mouse Brain Atlas(Fig. \ref{ABCDEFG1}E, F, and Figs. \ref{resolution}-\ref{xiaoshu_PAGA} in Appendix C). In particular, GATES can identify the narrow layer structure of MCL. If the interlamellar boundaries are indistinct or blurred, cells may exhibit abnormal morphologies or damage, indirectly affecting cardiac function~\cite{chronic}. In addition, the inner structure RMS that was blurred in the DAPI-stained image was clearly and unambiguously identified. Based on the successful identification and visualization of the overall spatial domains of the mouse olfactory bulb, we further focused our attention on the inner layers of the olfactory bulb. We observed with clarity the specific distribution patterns of cell types within the inner layers of the olfactory bulb. There is a distinct boundary between the cell types of the RMS and GCL. Specifically, granule cells are densely packed, whereas the GCL layer exhibits a more complex cellular structure, with some cells originating from the RMS layer interspersed within it (Fig. \ref{ABCDEFG1}G). This may indicate the existence of certain heterogeneities in morphology, function, and transcriptome between the cells of these two layers, or under certain physiological or pathological conditions, cells may migrate from their original positions to adjacent cellular layers. We further isolated and plotted each of the seven identified spatial domains into separate figures, allowing for more detailed observation and analysis of the characteristics of each spatial domain. Our method depicts the MCL layer as a continuous closed loop. This is consistent with the Allen Mouse Brain Atlas. Although STAGATE also identifies this layer, the recognition result is broader than the actual structure. Additionally, our method accurately identifies the RMS. In contrast, STAGATE exhibits fragmentation in identifying this region, while SCANPY fails to recognize it altogether. Similarly, SCANPY failed to identify the RMS and GCL, with their spots distributed across the entire region of the mouse olfactory bulb map. Notably, the ONL is situated at the outermost layer of the olfactory bulb, exhibiting a diversity in contour width across its various regions, with some areas barely discernible as thin lines. Despite this, our method is capable of accurately identifying the ONL without error. However, both STAGATE and SCANPY failed to fully delineate the entire contour of the ONL. Olfactory dysfunction may originate from structural abnormalities or functional impairments within the ONL. Our method enables precise localization of these abnormal regions. By comparing the characteristics of normal versus abnormal ONL layers, clinicians can more accurately assess the type and severity of olfactory disorders. (Fig. \ref{ABCDEFG1}H and Fig. \ref{detail_domain} in Appendix C). This further demonstrates the superiority of our method.

\subsubsection{The importance of gene expression similarity graph.} 
Finally, we performed an ablation experiment to compare the spatial domain recognition performance of the GATES method with/without gene expression similarity graph on the DLPFC dataset. We observed the GATES method was able to identify a certain spatial domain structure without incorporating the gene expression similarity graph. However, the consistency of the recognition results is poor. In contrast, when gene expression similarity graph is introduced, the spatial domain recognition ability of the GATES method is significantly improved. More importantly. It enhances the model's ability to decipher complex expression patterns, thus clear spatial domain boundaries can be stably delineated in all 12 slices, reflecting the real biological structure (Fig. \ref{ABCDEFG1}D). This significantly implies that by integrating gene expression data, we can gain a deeper understanding of the spatial heterogeneity within biological tissues, revealing the spatial distribution patterns of different cell types and their interactions.
\section{Discussion}
This paper presents a comprehensive graph construction strategy that integrates spatial proximity with gene expression similarity, allowing the network to learn both local and global information and thereby enhancing its ability to capture broader structural characteristics of biological tissues. This strategy challenges the assumption in previous methods that spatial proximity reflects gene expression similarity, which neglects non-local similarities between distant spots and leads to an overemphasis on local spatial information at the expense of global patterns. Specifically, our approach begins by constructing an expression graph that combines local and global information through spatial proximity and gene expression similarity. An autoencoder with adaptive attention is employed to assign appropriate weights to neighboring nodes, thereby enhancing feature extraction capabilities. 
By linearly combining spatial graph and expression graph, GATES effectively balances spatial context with gene expression data. 

Our method has been validated on DLPFC and mouse olfactory bulb datasets, demonstrating that integrating gene expression similarity for graph synthesis can significantly enhance spatial domain recognition performance. Notably, during the identification of complex domains within the mouse olfactory bulb, we observed that some cells from the RMS layer were dispersed within the GCL layer. This finding offers new perspectives and insights for investigating and understanding the mechanisms underlying disease pathogenesis.

In biological systems, the boundaries between different cellular layers are not distinct, rather they exist in a state of complex interactions and dynamic equilibrium. In practice, the precise identification of spatial domains and the extraction of spatially expressed genes are crucial for unraveling tissue organization and biological functions.
Our approach provides a more reliable tool for these areas, helping to improve the accuracy and efficiency of research and diagnosis. Naturally, the direction of the model can be further improved, and the composition can be carried out by integrating multimodal data, such as histological images and proteome data, to achieve more accurate detection of spatial domains and promote a more comprehensive exploration of the interior of biological tissues.

\section{Methods}
\subsection{Constructing graphs}
\subsubsection{Spatial adjacency graph.}

To combine adjacent spot similarities, GATES constructs a spatial neighbor graph \( G^{\rm spatial}(V, E) \) based on a predefined radius \(r\) and spatial location information, selecting 6 neighboring spots for each defined spot. Let \(\mathbf{A}\) be the spatial adjacency matrix of size \(\left | V \right | \times \left | V \right |\), where \(\mathbf{A}(i,j)=1\) indicates that vertices \(i\) and \(j\) are adjacent, meaning their Euclidean distance is less than \(r\). To account for self-loops, we augment \(\mathbf{A}\) with the identity matrix \(\mathbf{I}\), forming \(\mathbf{A}^{*}= \mathbf{I}+ \mathbf{A}\), which ensures both neighbor connections and self-characteristics are captured.

\subsubsection{Gene expression graph.}
To capture global information, GATES constructs an undirected neighbor graph \( G^{\rm gene}(V, E) \) based on gene expression similarities. For each spot, we compute its gene expression similarity with all other spots (e.g., using cosine similarity) and select the \( k \) spots with the highest similarity scores as nearest neighbors, forming the gene expression similarity adjacency matrix \( \mathbf{B} \). The matrix \( \mathbf{B} \), of size \( |V| \times |V| \), has \(\mathbf{B}(i, j) = 1\) if the similarity score between spots \( i \) and \( j \) is within the top \( k \), indicating an edge between vertices \( i \) and \( j \). To account for self-loops, we augment the matrix as \( \mathbf{B}^* = \mathbf{I} + \mathbf{B} \), where \( \mathbf{I} \) is the identity matrix, representing the gene expression graph with self-connections.

\subsection{Graph attention auto-encoder}
Graph autoencoder is an unsupervised learning technique that excels at uncovering latent structures and key features by learning low-dimensional representations. Integrating an attention mechanism allows adaptive weighting of neighboring vertices, improving feature extraction.

\subsubsection{Encoder.}
Let \( L \) denote the number of layers in the encoder and \( S \) represent the number of spots in the slice. The encoder primarily extracts the low-dimensional embedding representation of the input graph structure as \( d \). In layer \( k \) (\( k \in \{1, 2, \ldots, L-1\} \)), the \( k \)-th encoder layer generates the embedding for spot \( i \) (\( i \in \{1, 2, \ldots, S\} \)) as follows: 
\begin{equation}
    \mathbf{d}_{i}^{(k) } = \sigma \left (  {\textstyle \sum_{j\in N_{i} }^{}}{\rm \mathbf{att}}_{ij}^{(k)} \left ( \mathbf{W}_{k} \mathbf{d}_{j}^{(k-1)}   \right )   \right ) ,
\end{equation}
where $\textbf{W}_{k} $ is the trainable weight matrix of encoder, $N_i$ is the set of neighbors for spot $i$, $\sigma$ is the nonlinear activation function, and ${\rm \mathbf{att}}_{ij}^{\left(k \right)}$ is the edge weight of the normalized edge between the spot $i$ and the spot $j$ in the graph attention layer. In our method, the $L$-th encoder layer operates without the attention mechanism and is expressed as follows:
\begin{equation}
    \mathbf{d}_{i}^{(L)}=\sigma\left(\mathbf{W}_{L} \mathbf{d}_{i}^{(L-1)}\right). \\
\end{equation}
The output of the final layer of the encoder is regarded as the ultimate embedding.

\subsubsection{Decoder.} 
The decoder is responsible for reconstructing the latent representation into the original expression data. 
The embedding of the $k$-th layer decoder reconstruction spot $i$ in the $(k-1)$-th layer is defined as:
\begin{equation}
    \hat{\mathbf{d}} _{i}^{k-1}=\sigma\left( {\textstyle \sum_{j\in N_{i} }^{}}  \hat{\mathbf{att}}_{ij}^{(k-1)}\left(\hat{\mathbf{W}}_{k} \hat{\mathbf{d}}_{j}^{(k)}\right)\right). \\
\end{equation}
where $\hat{\mathbf{att}}_{ij}$ represents the edge weight of the decoder's attention layer, and $\hat{\mathbf{W}}_{k}$ is the decoder's trainable weight matrix.
Analogous to the encoders, the final layer forgoes the utilization of the attention mechanism:
\begin{equation}
    \hat{\textbf{d}}_{i}^{\left(0\right)}  =\sigma \left ( \hat{\textbf{W}}_{1}\hat{\textbf{d}}_{i}^{\left ( 1 \right ) }     \right ). 
\end{equation}
%The output of the decoder is considered to be a normalized expression of the reconstruction.
To mitigate overfitting in our method, we separately set $\hat{\textbf{W}}_{k} = \textbf{W}_{k}{^{T}}$ and $\hat{\textbf{att}}^{(k)} = \textbf{att}^{(k)}$.

\subsubsection{Graph attention layer.}
The attention mechanism operates as a single-layer feedforward neural network, with shared parameters among nodes that are defined by weight vectors. In the \( k \)-th encoder layer, the calculation of the edge weight from node \( i \) to its adjacent neighbor \( j \) is expressed as follows:
\begin{equation}
    e_{ij}^{(k)}=\operatorname{Sigmoid}\left(\mathbf{v}_{s}^{(k)^{T}}\left(\mathbf{W}_{k} \mathbf{d}_{i}^{(k-1)}\right)+\mathbf{v}_{r}^{(k)^{T}}\left(\mathbf{W}_{k} \mathbf{d}_{j}^{(k-1)}\right)\right),
\end{equation}
where $\mathbf{v}_{s}^{(k)}$ and $\mathbf{v}_{r}^{(k)}$ are the trainable weight vectors in the $k$-th layer, applied to the transformed embeddings of spot $i$ (source) and spot $j$ (receiver) for distinct operations.
%where $\mathbf v_{s}^{(k)}$ and $\mathbf v_{r}^{(k)}$ are the trainable weight vectors of the $k$-th layer. 
To facilitate the comparison of coefficients among nodes, we normalize the selections for \( j \):
\begin{equation}
    {\rm att}_{i j}^{(k)}=\frac{\exp\left ( e_{ij}^{(k)}  \right ) }{\sum_{j\in N_{i} }\exp\left ( e_{ij}^{(k)}  \right )  }. 
\end{equation}

GATES employs a self-attention mechanism for two distinct types of networks: spatial similarity and gene expression similarity. It linearly combines the information learned at each layer from both networks to obtain the final information for each layer:
\begin{equation}
   \mathbf{att}_{ij}=\alpha \mathbf{att}_{ij}^{\rm spatial}+\left ( 1-\alpha  \right ) \mathbf{att}_{ij}^{\rm gene}, 
\end{equation}
where \(\mathbf{att}_{ij}\) denotes the edge weight between spots \(i\) and \(j\) in the output of the fused attention layer, while \(\mathbf{att}_{ij}^{\rm spatial}\) and \(\mathbf{att}_{ij}^{\rm gene}\) represent the edge weights between spots \(i\) and \(j\) in the outputs of the learned spatial adjacency graph and the gene expression graph, respectively. $\alpha$ is a hyperparameter that controls the linear combination of these features, enabling GATES to effectively balance spatial context and gene expression data through the fusion of both types of information.

\subsection{Loss function}
The objective function of the entire model framework is to maximize the similarity between the input expression and the reconstructed expression by minimizing the mean square error loss $L_{\rm recon}$:
\begin{equation}
    L_{\rm recon}  =  {\textstyle \sum_{i=1}^{S}}  \left \| \textbf{x}_{i}-\hat{\textbf{d}}_{i}^{(0)} \right \|_{2}.
\end{equation}

\bibliographystyle{ieeetr} 
\bibliography{ref}

\newpage

\appendix
\section{Experimental Details}
\subsection{Data description}
We performed spatial domain detection on the DLPFC dataset generated by the 10x Visium platform and the mouse olfactory bulb dataset generated by the Stereo-seq platform.
The DLPFC dataset encompasses twelve distinct tissue sections sourced from three adult specimens. Each is composed of four consecutive slices, individually displaying a range of five to seven layers of DLPFC and white matter (WM). It can be accessed online via the platform \url{http://research.libd.org/spatialLIBD/}.
The mouse olfactory bulb dataset has a higher resolution than the DLPEC dataset. It can be accessed online via the platform \url{https://db.cngb.org/search/project/CNP0001543/}.

\subsection{Code Availability}
The GATES algorithm is implemented in Python and is accessible on GitHub at \url{https://github.com/xiaoxiongtao/GATES}.

\subsection{Data preprocessing}
First, we exclude data spots that represent major organizational areas to reduce noise and interference, ensuring focused and accurate analysis. This was followed by logarithmic transformation and normalization to enhance the linearity and stability of the data, as well as to eliminate bias due to different sequencing depths. In the end, we selected the top 3000 genes with the highest variability to reduce the complexity of the analysis. At the same time, we focused on those genes that are most likely to reveal biological patterns.

\subsection{Evaluation indicators}
We use the Adjusted Rand Index (ARI) as an important measure of the similarity between the two regions. It evaluates the spatial domain recognition performance by considering the distribution of data spot pairs in different clustering results. The range of the adjusted Rand index is $\left [ -1,1 \right ]$. The value of ARI is $1$ when the results of the two clusters are the same, close to 0 when the results of the two clusters are completely random and close to $-1$ when the results of the two clusters are completely different. The formula is as follows:
\begin{equation}
   {\rm ARI}=\frac{\sum_{i j}\binom{n_{i j}}{2}-\left[\sum_{i}\binom{a_{i}}{2} \cdot \sum_{j}\binom{b_{j}}{2}\right] /\binom{n}{2}}{\frac{1}{2}\left[\sum_{i}\binom{a_{i}}{2}+\sum_{j}\binom{b_{j}}{2}\right]-\left[\sum_{i}\binom{a_{i}}{2} \cdot \sum_{j}\binom{b_{j}}{2}\right] /\binom{n}{2}}
\end{equation}
where $n$ is the number of spots and $ n_{ij} $ is the number of samples with true label $i$ and cluster label $j$.

\newpage
\section{Experiments on DLPFC Dataset}

\begin{figure}[!htb]
\centering
\includegraphics[width=0.9\textwidth]{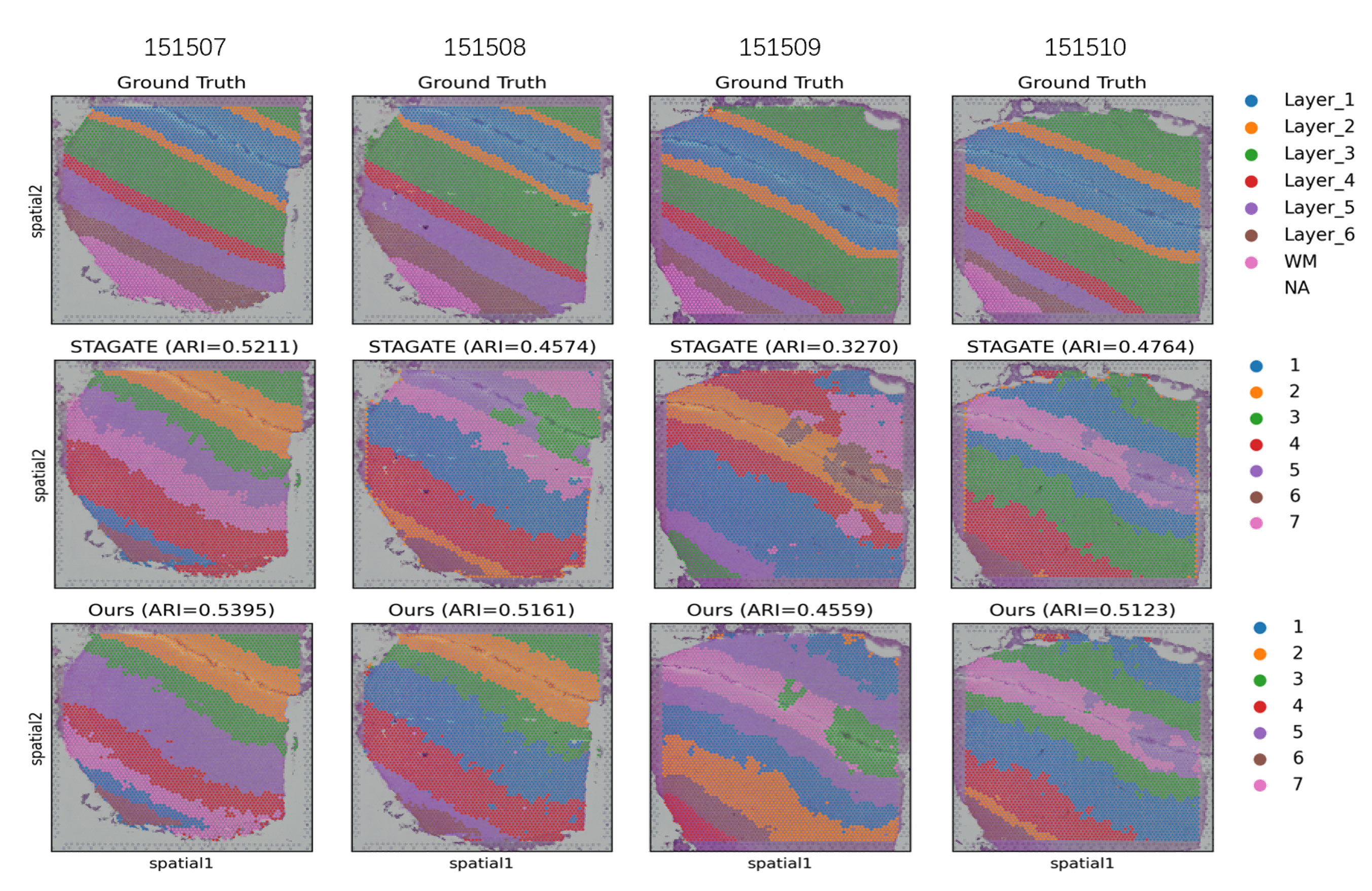}
\caption{Visualization of 151507-151510 slice spatial domain recognition results.} \label{domain1}
\end{figure}

\begin{figure}[!htb]
\centering
\includegraphics[width=0.9\textwidth]{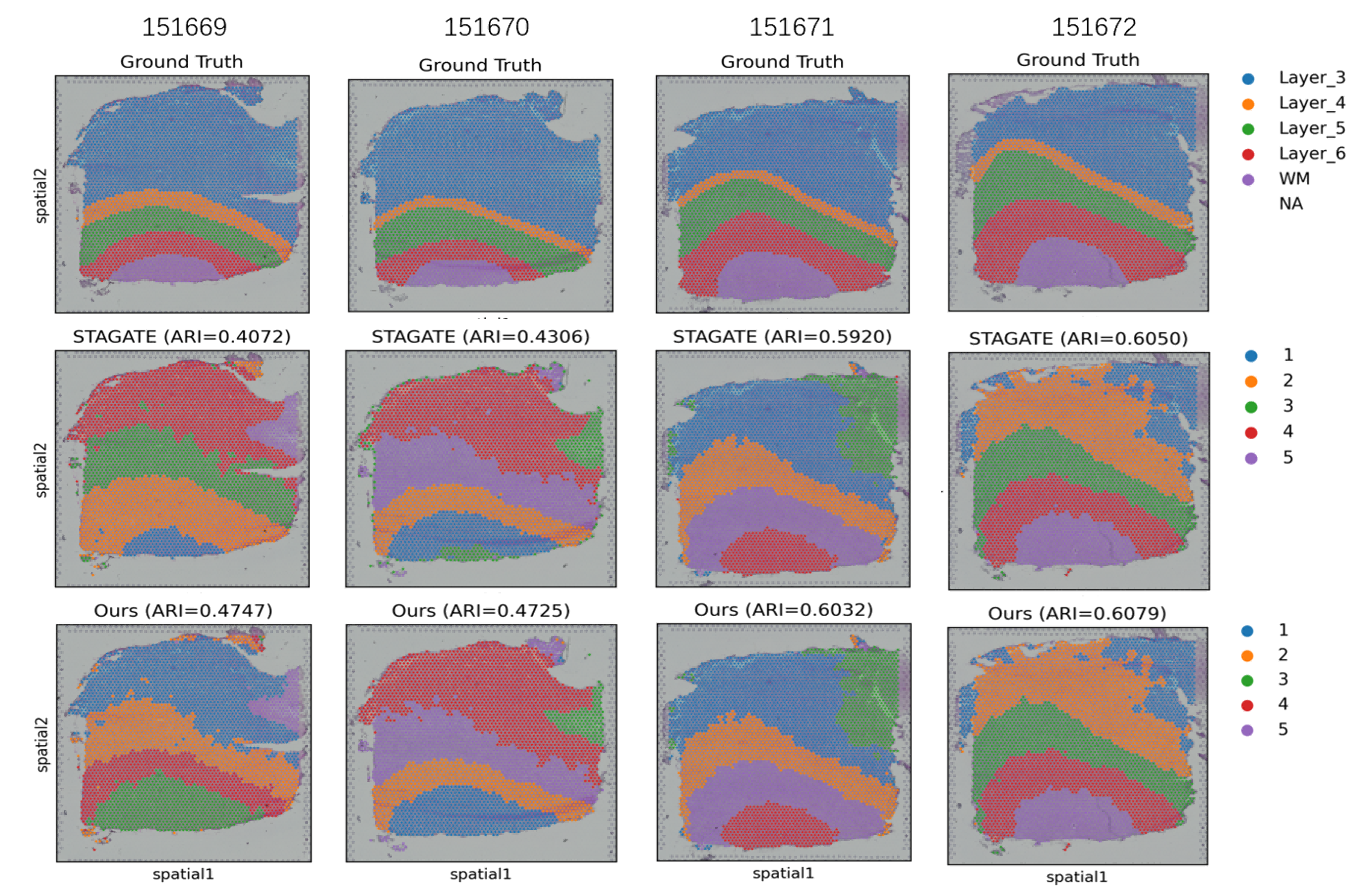}
\caption{Visualization of 151669-151670 slice spatial domain recognition results.} \label{domain2}
\end{figure}

\newpage
\begin{figure}[!htb]
\centering
\includegraphics[width=0.9\textwidth]{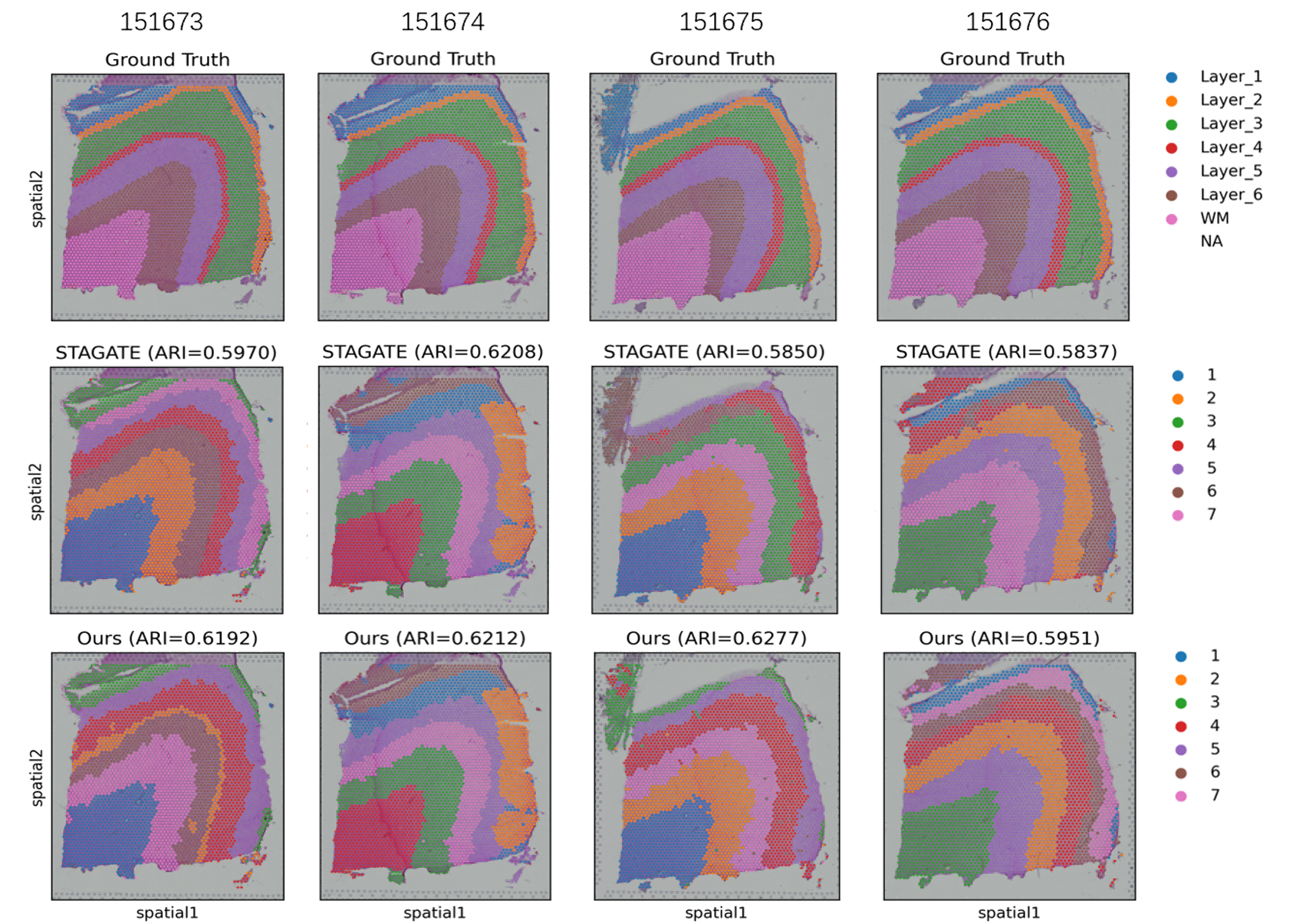}
\caption{Visualization of 151673-151676 slice spatial domain recognition results.} \label{domain3}
\end{figure}

\begin{figure}[!htb]
\centering
\includegraphics[width=\textwidth]{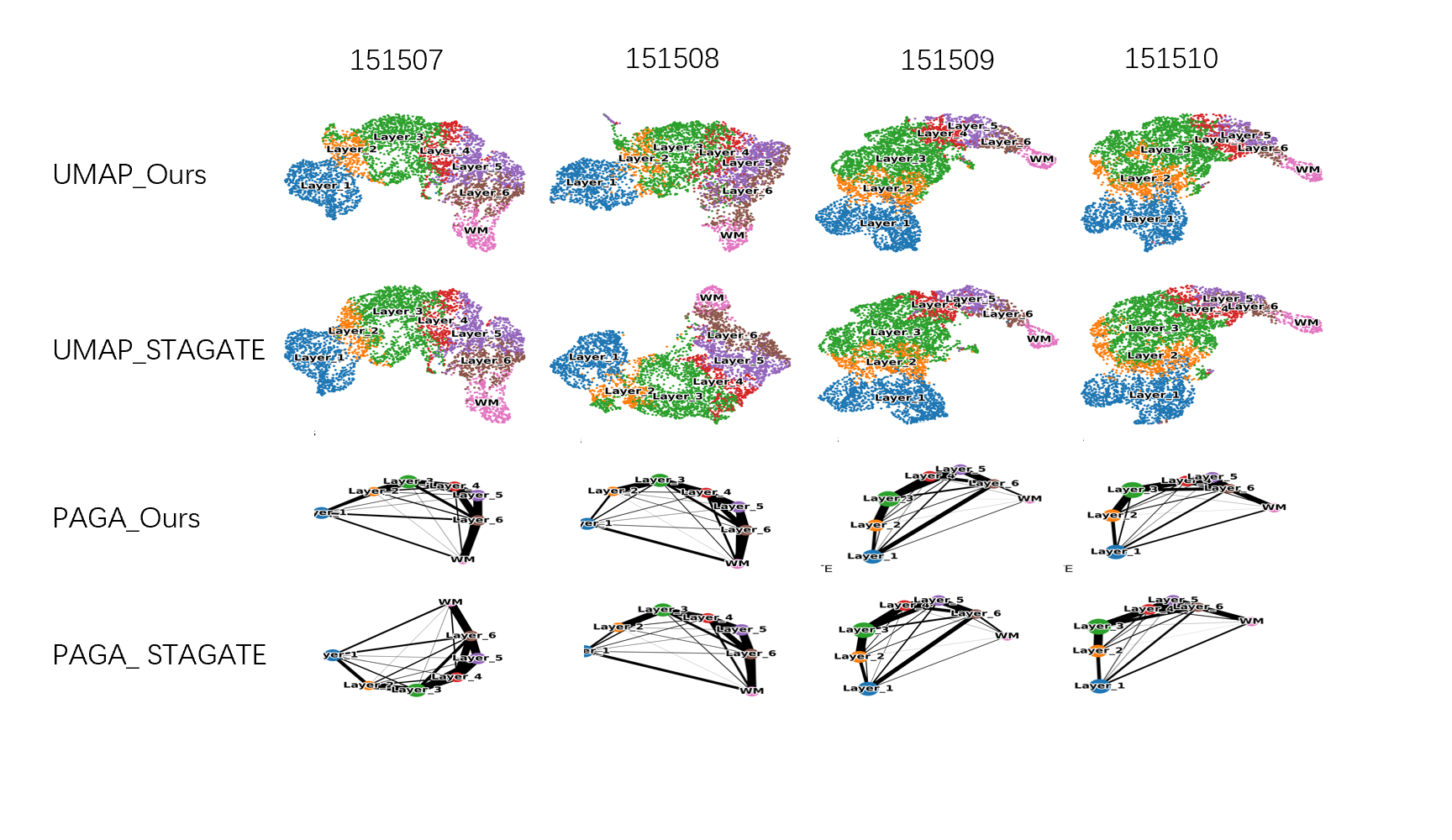}
\caption{UMAP and trajectory inference results for 151507-151510 slices.} \label{PAGA1}
\end{figure}

\begin{figure}[htbp]
\includegraphics[width=\textwidth]{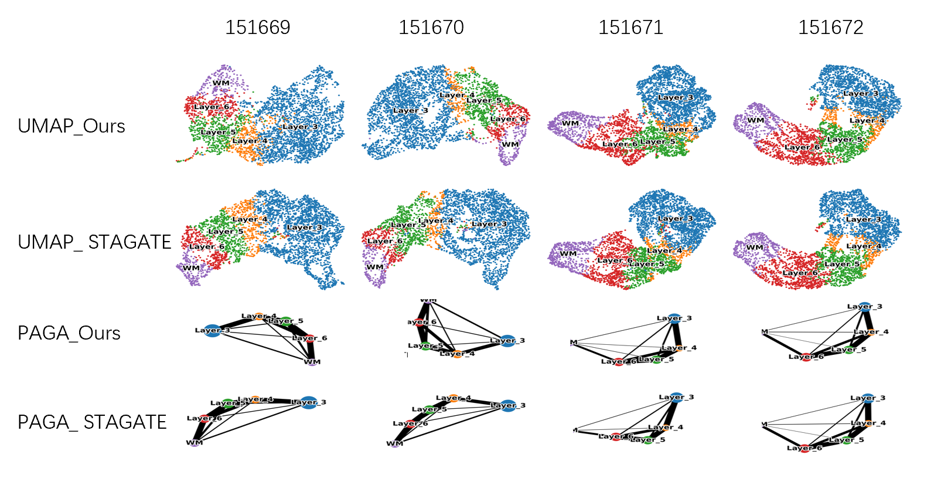}
\caption{UMAP and trajectory inference results for 151507-151510 slices.} \label{PAGA2}
\end{figure}

\begin{figure}[htbp]
\centering
\includegraphics[width=\textwidth]{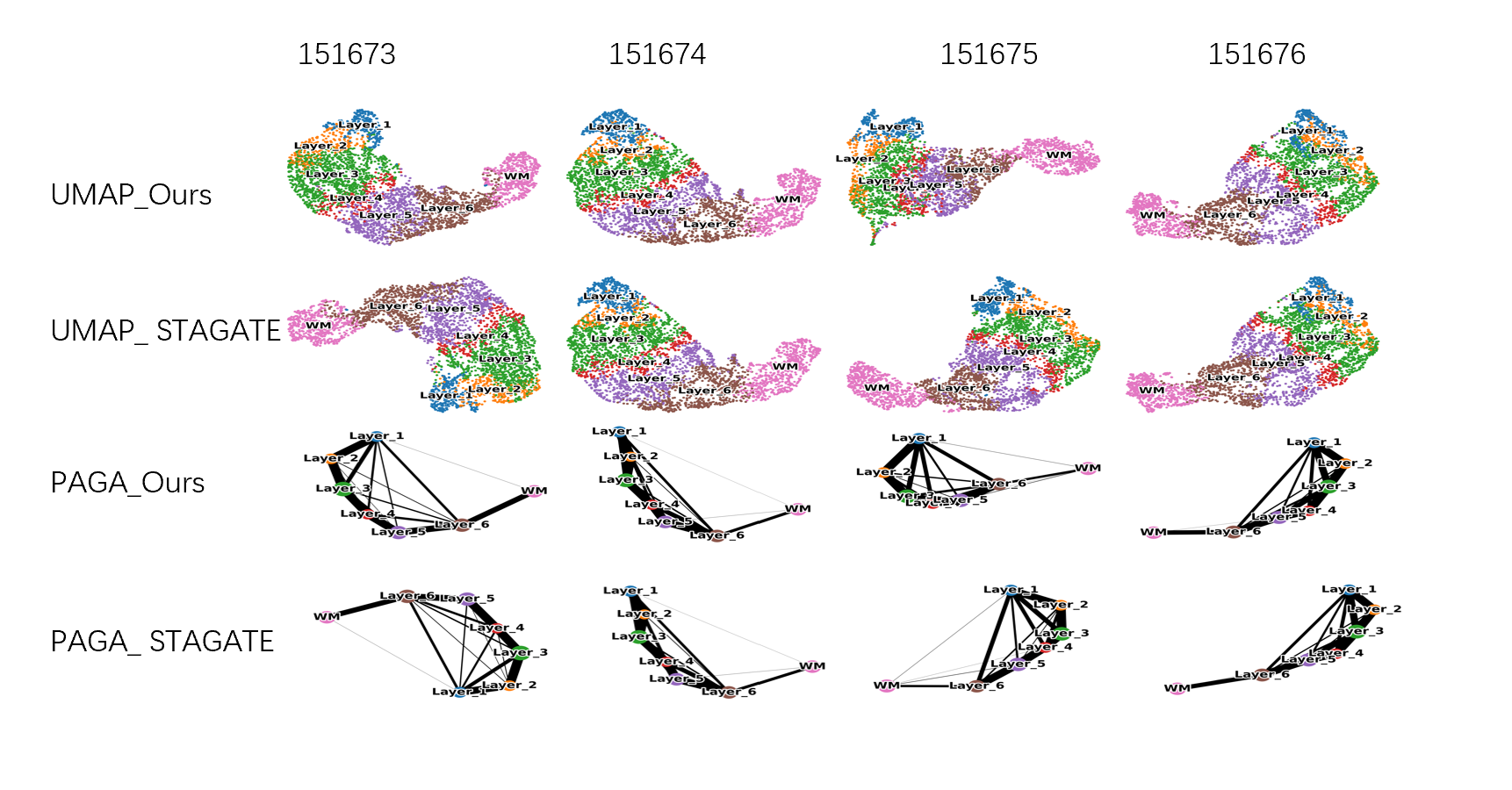}
\caption{UMAP and trajectory inference results for 151507-151510 slices.} \label{PAGA3}
\end{figure}

\newpage

\section{Experiments on Mouse Olfactory Bulb Dataset}
\begin{figure}[!htb]
\centering
\includegraphics[width=0.9\textwidth]{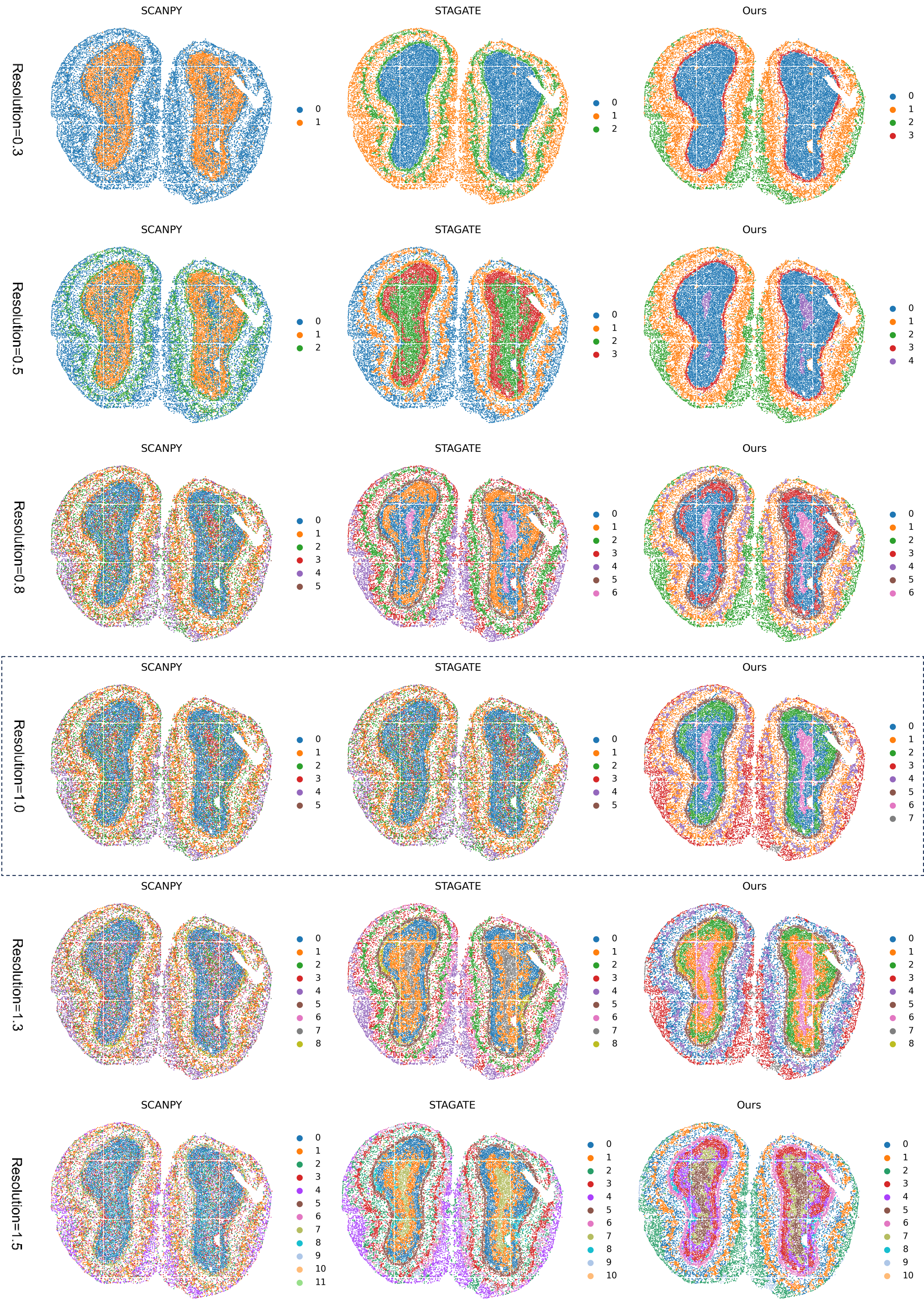}
\caption{Spatial domain recognition results of mouse olfactory bulb tissue dataset with different precisions. In the absence of prior information, we manually selected the resolution using the Louvain algorithm, showcasing its results at various resolutions. We ultimately chose a resolution of 1 as the final parameter for our method.} \label{resolution}
\end{figure}

\begin{figure}[!htb]
\includegraphics[width=0.9\textwidth]{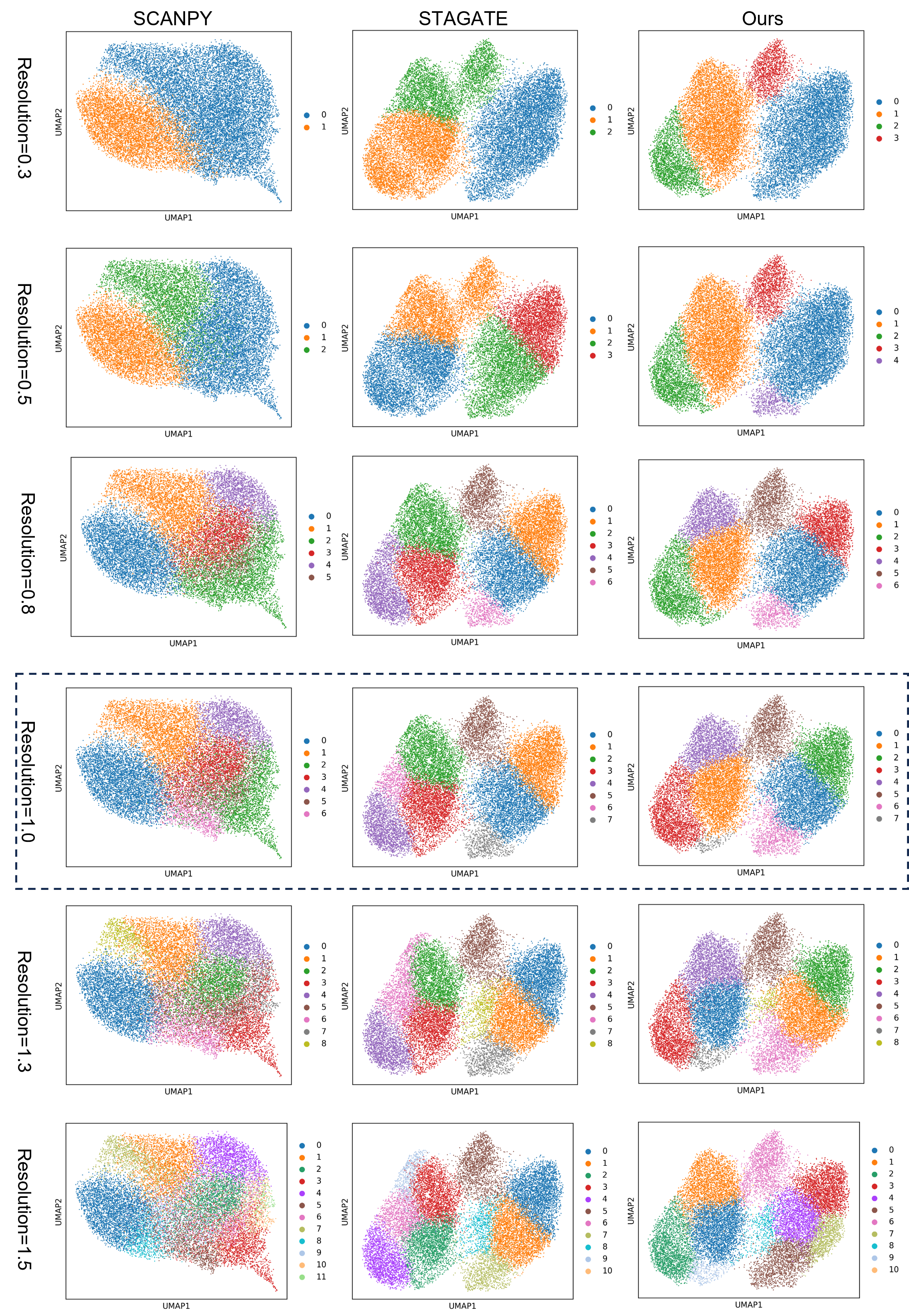}
\caption{UMAP in the mouse olfactory bulb tissues dataset with different precisions. We ultimately chose a resolution of 1 as the final parameter for our method. 
} \label{xiaoshu_UMAP}
\end{figure}

\begin{figure}
\includegraphics[width=\textwidth]{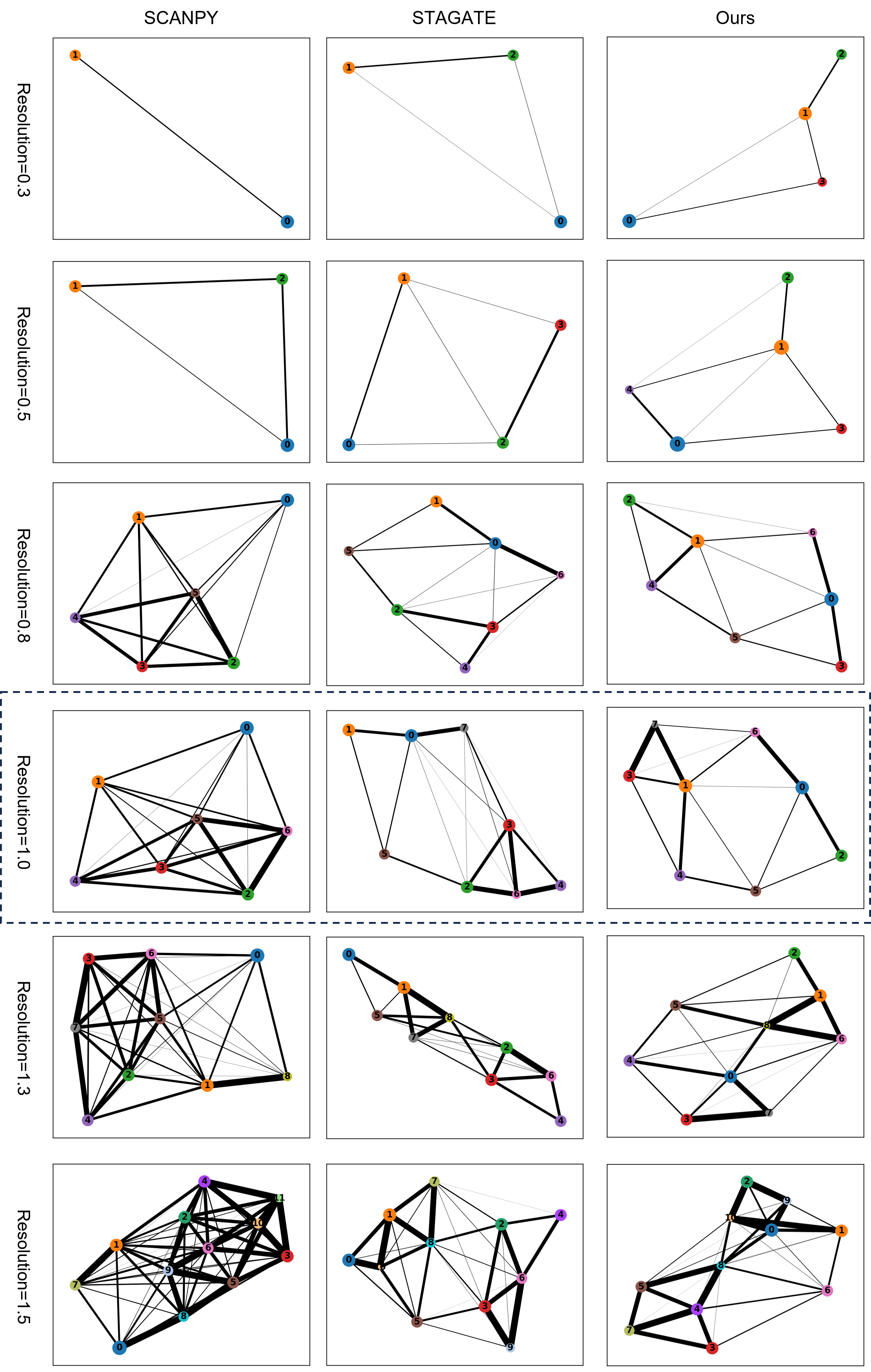}
\caption{trajectory inference in the mouse olfactory bulb tissues dataset with different precisions. We ultimately chose a resolution of 1 as the final parameter for our method.} \label{xiaoshu_PAGA}
\end{figure}

\begin{figure}
\centering
\includegraphics[width=0.9\textwidth]{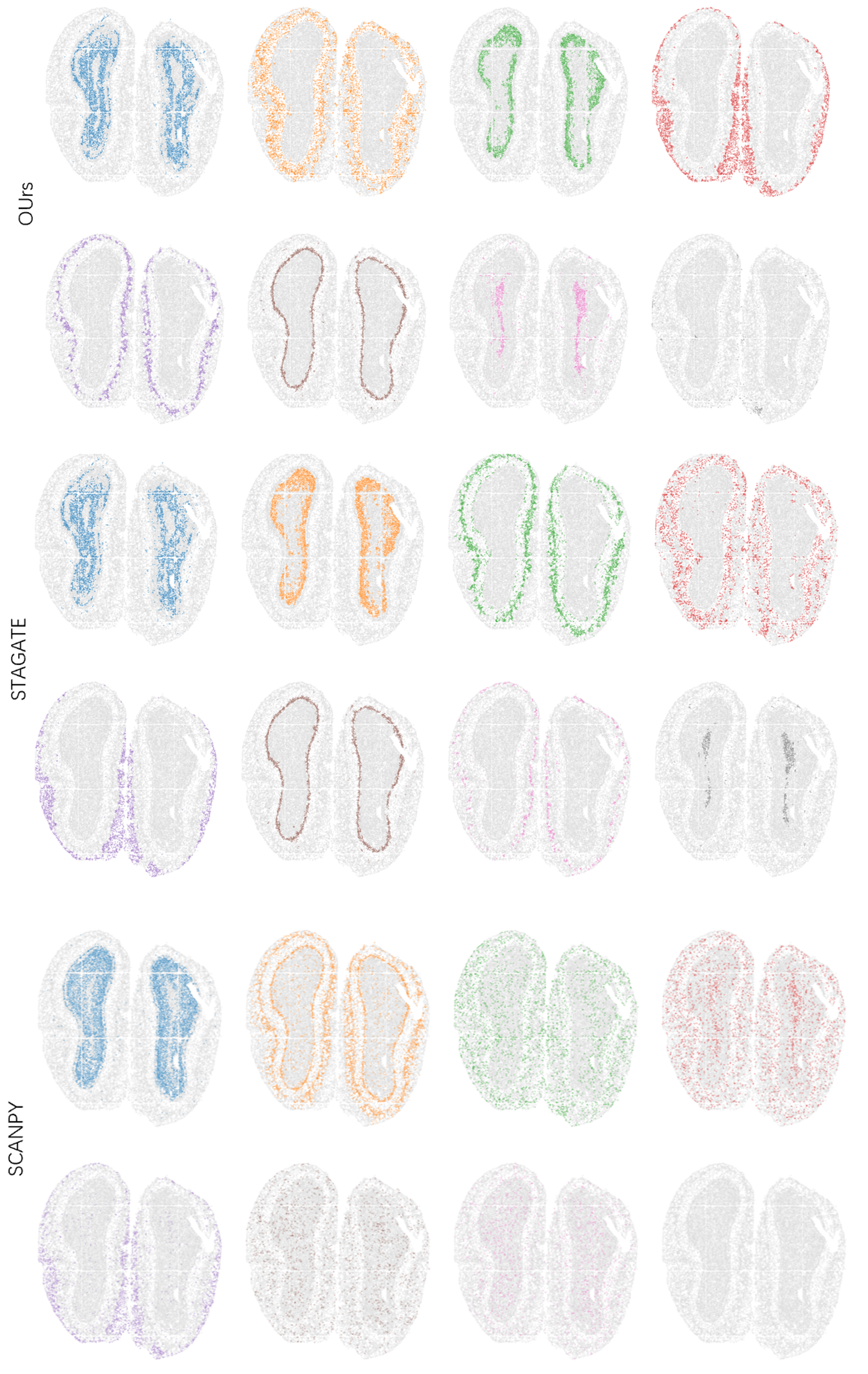}
\caption{The detailed diagram illustrates the spatial domains identified by various methods. SCANPY mixes spots from different clusters and struggles to accurately identify domains like RMS, GCL, and IPL, highlighting its limitations. In contrast, GATES and STAGATE consider spatial locations, successfully identifying several layers of the mouse olfactory bulb. However, STAGATE struggles with clear delineation of the ONL, IPL, and RMS layers, resulting in fragmentation and a broader appearance. GATES outperforms by integrating global gene expression data, providing a clearer boundary for the GL layer that aligns more closely with its actual size compared to STAGATE.} \label{detail_domain}
\end{figure}
\end{document}